\documentclass[prb,amsmath,twocolumn,showpacs,superscriptaddress]{revtex4-1}
\usepackage{array}
\usepackage[titletoc,title]{appendix}
\usepackage{booktabs}
\usepackage{float}
\usepackage{tabularx}
\usepackage{color}
\usepackage{natbib}
\usepackage{wrapfig}
\usepackage{graphicx}
\usepackage{psfrag}
\usepackage{multirow}
\usepackage{subfig}	
\usepackage[english]{babel}
\usepackage{amsmath,amsthm,amssymb,amsfonts,latexsym} 
\usepackage[latin1]{inputenc} 
\pdfpageattr {/Group << /S /Transparency /I true /CS /DeviceRGB>>}  
\newlength{\picwidth}
\newcommand{\ra}[1]{\renewcommand{\arraystretch}{#1}}

\begin{document}

\title{Van Hove singularity and ferromagnetic instability in phosphorene}
\author{A. Ziletti}
\thanks{A. Ziletti and S. M. Huang contributed equally to this work.}
\affiliation{Department of Chemistry, Boston University, 590 Commonwealth Avenue, Boston Massachusetts 02215, USA}

\author{S. M. Huang}
\thanks{A. Ziletti and S. M. Huang contributed equally to this work.}
\affiliation{Centre for Advanced 2D Materials and Graphene Research Centre, National University of Singapore, Singapore 117546}
\affiliation{Department of Physics, National University of Singapore, Singapore 117542}

\author{D. F. Coker}
\affiliation{Department of Chemistry, Boston University, 590 Commonwealth Avenue, Boston Massachusetts 02215, USA}

\author{H. Lin}
\email{nilnish@gmail.com}
\affiliation{Centre for Advanced 2D Materials and Graphene Research Centre, National University of Singapore, Singapore 117546}
\affiliation{Department of Physics, National University of Singapore, Singapore 117542}

\begin{abstract}
Using Wannier function-based interpolation techniques, we present compelling numerical evidence for the presence of a saddle-point van Hove singularity at the $\Gamma $ point near the phosphorene Fermi energy. We show that in proximity of the van Hove singularity the spin susceptibility presents the logarithmic temperature dependence typical of Liftshitz phase transitions. Furthermore, we demonstrate that the critical temperature for the ferromagnetic transition can be greatly increased (up to 0.05~K) if strain along the zigzag ridges is applied. 
Although the ferromagnetic state would be very difficult to experimentally reach, the logarithmic temperature behaviour of the spin susceptibility due to the van Hove singularity is found to persist at much higher temperatures (up to $\sim$97~K).
\end{abstract}
\maketitle


\section{Introduction}

Saddle-point van Hove singularities\cite{vanhove1953} (VHSs) originate from saddle points in the band structure, around which the band curvature has opposite signs along two orthogonal directions.
In two dimensions, the density of states (DOS) diverges at the VHS, and therefore arbitrary weak interactions can produce large effects in the electronic behaviour, giving rise to instabilities in many aspects such as charge, spin, and/or pairing susceptibilities. 
Once the Fermi energy approaches a VHS, ferromagnetism,\cite{MFleck1997,Hlubina1997} antiferromagnetism,\cite{HQLin1987} and/or superconductivity\cite{WKohn1965,Hirsch1986,Honerkamp2001}  can be substantially enhanced.

The VHS is a topological critical point of the Fermi surface,
across which the quantum Lifshitz phase transition takes place.\cite{Lifshitz,Yamaji2006,NSVidhyadhiraja2009,KSChen2011}
The Lifshitz transition for noninteracting systems is continuous and does not break symmetry.
For interacting systems, however, the Lifshitz transition may become discontinuous and accompany symmetry breaking. \cite{Yamaji2006,Okamoto2010}
In cuprates, Hall coefficient measurements provide evidence for the Fermi surface topology change. \cite{Jiang1994,Dagan2004} The Lifshitz transition is also proposed to change the Fermi liquid into the marginal Fermi liquid,\cite{Chen2012} and the VHS is thus argued to be responsible for the linear temperature ($T$) dependence of resistivity and the $T$-independent thermopower\cite{DMNewns1994} observed in this regime.\cite{SDObertelli1992,JLCohn1991,BFisher1993}
Moreover, in the so-called ``van Hove scenario", the presence of a VHS near the Fermi energy is argued to play a major role in the high-$T_c$ superconductivity of cuprates.\cite{DMNewns1992,RSMarkiewicz1997} 
Given the strong influence of VHSs on the properties of materials, it is important to identify the presence and understand the role of these singularities, especially for technologically promising low-dimensional materials like phosphorene.

Phosphorene,\cite{liu2014,rodin2014} a single layer of black phosphorus, is the most recent addition to the growing family of two dimensional (2D) materials. It is a semiconductor with high potential for applications in electronic and optoelectronic devices.\cite{buscema2014} Despite the relative infancy of the field, few-layer phosphorene field effect transistors exhibit very high on-off current ratios\cite{li2014,SPKoenig2014} (exceeding 10$^5$) and ambipolar behaviour,\cite{RADoganov2014} together with the highest hole mobility ever (4000~cm$^2$/Vs) for a 2D material apart from graphene.\cite{NGillgren2014} 
Phosphorene's pliable waved structure also allows for strain engineering of both effective masses and bandgaps.\cite{JWJiang2014} Strain can even induce a semiconductor to metal transition.\cite{rodin2014} 

In this article, we show that a VHS is present at the phosphorene Fermi energy, and we investigate the consequent ferromagnetic instability in both the unstrained and strained cases.

The article is organized as follows: after introducing the computational methodology, in Sec.  \ref{subsec:van-hove} we present the electronic structure of phosphorene near the VHS, in Sec. \ref{subsec:magnetism} we study the ferromagnetic instability (without strain), and finally in Sec. \ref{subsec:strain} we investigate the effect of strain on the critical temperature $T_c$ of the ferromagnetic transition.


\section{Computational details}
\label{sec:comp-details}

The calculations involve the following three steps. \\
(i) A density functional theory (DFT) calculation is performed with a plane wave basis set, as implemented in the {\sc Quantum ESPRESSO} package.\cite{qe} We use the PBEsol functional\cite{PBEsol} for the exchange and correlation energy. A plane wave basis set with a kinetic energy cutoff of 70~Ry (280~Ry) is used to represent the electronic wave function (charge density). The core electrons are described via the projected-augmented wave (PAW)\cite{paw} method; 12.9 \AA\ of vacuum are added in the direction normal to the monolayer to avoid spurious interactions between periodic replicas. Both lattice parameters and atomic positions are relaxed until the forces on each atom are less than 10$^{-3}$~eV/\AA\ and the pressure is less than 1~kbar. After lattice relaxation, the phosphorene crystal parameters are $a_x$=3.28~\AA\ and $a_y$=4.44~\AA, in agreement with a previous study.\cite{rodin2014} The optimized configuration of the phosphorene monolayer is presented in Fig. \ref{fig:1}(a).
In this DFT calculation, the Brillouin Zone (BZ) is sampled using a $\Gamma $-centered 60$\times$48$\times$1 Monkhorst-Pack (MP) grid.\cite{mpgrid}
This calculation will serve as a benchmark for the Wannier interpolation of the band structure [Fig. \ref{fig:1}(b) and Fig. \ref{fig:1}(c)].

(ii) Using the self-consistent charge density obtained from step (i), we evaluate the required input quantities for the Wannier calculation (energy eigenvalues, overlap matrices and projections\cite{marzari1997}) on a relatively coarse 10$\times$8$\times$1 mesh for the unstrained case (Secs. \ref{subsec:van-hove} and  \ref{subsec:magnetism}), and a 30$\times$8$\times$1 mesh for the strained case (Sec. \ref{subsec:strain}). These $k$-point meshes are fine enough to provide converged Wannier functions. The calculations are performed with {\sc Quantum ESPRESSO} and its post-processing subroutine {\sc pw2wannier90}.

The aim of the two previous steps is to obtain the maximally localized Wannier functions\cite{marzari1997,marzari2012} (MLWFs) to be later used for the very dense $k$-point sampling around the VHS.

(iii) With the energy eigenvalues, overlap matrices and projections obtained from step (ii), we construct the MLWFs according to the procedure presented in Ref. \onlinecite{marzari1997} and Ref. \onlinecite{souza2001}. The resulting Wannier functions consist of three $p$-orbitals centered on each P atom, leading to the wannierization of 6 valence and 6 conduction bands (there are four P atoms in the phosphorene unit cell).

One of the main advantages of the maximally localized Wannier representation of the DFT orbitals is that quantities calculated on a coarse reciprocal-space grid can be used to interpolate on a much finer grid with low computational cost. The Wannier interpolation is particularly useful when a fine BZ sampling is required to converge the quantity of interest. In this work, such quantities are the DOS and the valence band in a small region around the VHS.
For the DOS calculation, an extremely dense Wannier interpolated mesh of 4$\cdot$$10^{4}$$\times$3.2$\cdot$$10^{4}$$\times$1 - corresponding to $\sim$1.3 billion $k$-points in the BZ - is used to capture the sharp peak in the DOS due to the VHS. We use a smearing of 7$\times$10$^{-4}$~eV.
Similarly, a Wannier interpolation with a reciprocal lattice spacing of $\Delta k_{x,y}$=2$\cdot10^{-3}\times 2\pi/a_{x,y}$ corresponding to a 500$\times$500$\times$1 MP grid is employed for the contour plot of the valence band around the VHS.
All MLWF calculations are performed with the {\sc Wannier90} package.\cite{wannier90}

To further validate the PBEsol results, we have performed additional calculations with the local LDA functional,\cite{lda} the semilocal PBE functional\cite{pbe}, the screened-hybrid HSE06 functional,\cite{hse06} and the $GW$ method\cite{hybertsen1986,hedin1965} in order to elucidate the position of the valence band maximum. In particular, the HSE06 functional and the $GW$ method are known to provide a more accurate description of the electronic properties of semiconductors and insulators than local or semilocal density functionals.\cite{janesko2009,fuchs2007}

In LDA and PBE calculations, atomic positions and lattice parameters are relaxed till the forces are less than 10$^{-3}$ eV/\AA\ and the pressure less than 1~kbar. The BZ is sampled with a $\Gamma$-centered 60$\times$48$\times$1 MP grid as in the case of PBEsol calculations.
The HSE06 corrections are instead calculated self-consistently using the PBE relaxed lattice parameters and atomic positions, together with a 12$\times$12$\times$1 MP grid. The band structure is then obtained using the derived Wannier functions in a similar fashion to the PBEsol calculations outlined above. 

The $GW$ calculations are performed in two steps. First, atomic positions and relaxed lattice geometries are calculated with the PBE functional and norm-conserving Troullier-Martins pseudopotentials.\cite{mt-pseudo} Then the $GW$ corrections are computed following the method proposed by Hybertsen and Louie.\cite{hybertsen1986} We include 138 bands in the evaluation of the dielectric matrix and the self-energy, with a cutoff of 4~Ry for the dielectric matrix; convergence was checked including up to 384 bands. A supercell of 20~\AA\ in the direction perpendicular to the monolayer and a slab-truncation potential\cite{ismail-beigi2006} are used in these $GW$ calculations to avoid spurious interactions with periodic replicas of the system. A fine MP grid is employed in the direction of the VHS (100$\times$8$\times$1) to distinguish the top of the valence band from the $\Gamma$ point, thus providing evidence for the presence of the VHS.  All $GW$ calculations are performed with the plane-wave based ABINIT package.\cite{abinit}


\section{Results and discussion}

\subsection{Electronic band structure and van Hove singularity}
\label{subsec:van-hove}

\begin{figure}[htb]
\centering
    \includegraphics*[trim=0pt 0pt 0pt 0pt, width=8.6cm]{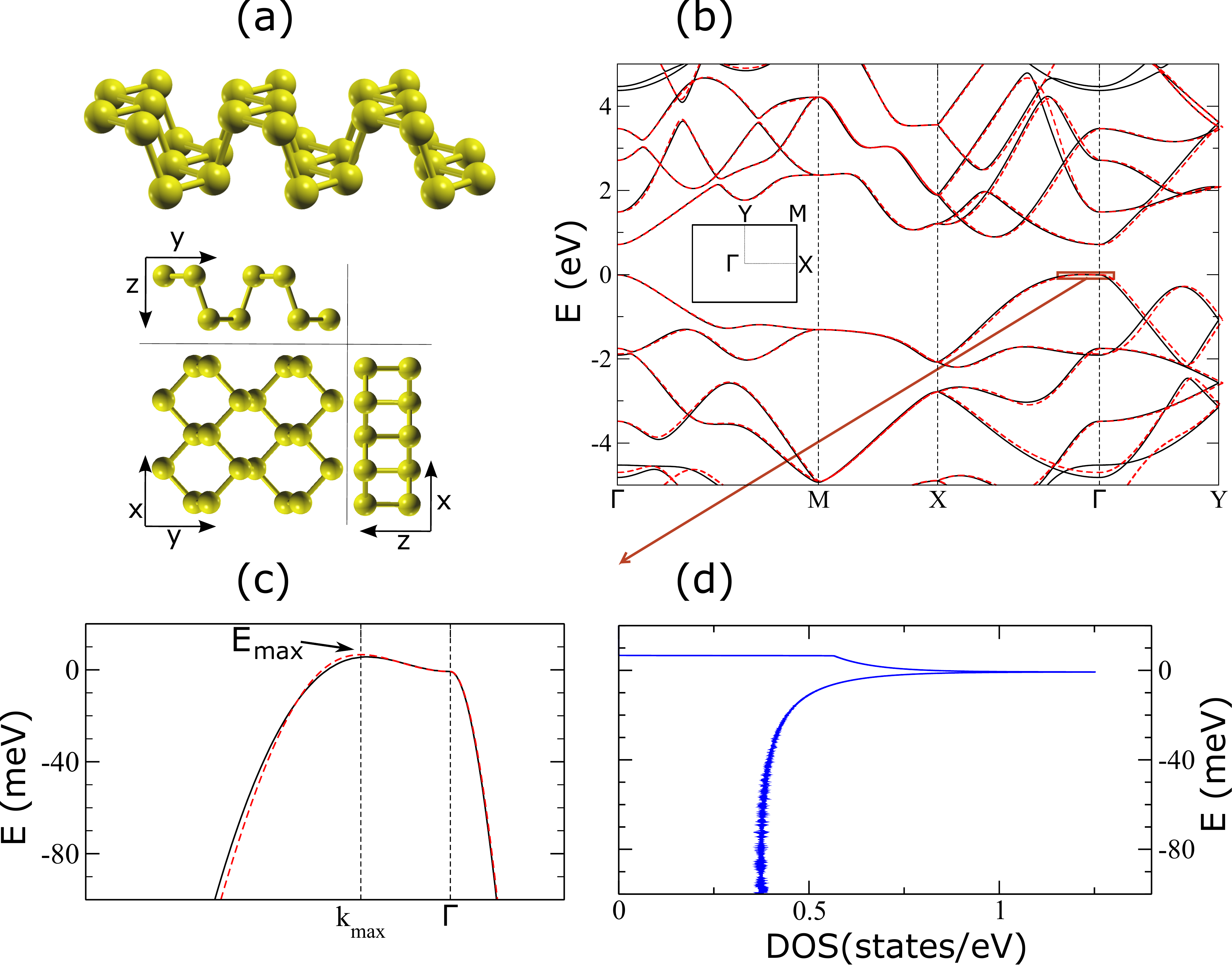}
 \caption{\small (Color online) (a) Crystal structure of phosphorene and its projections on the $y$-$z$, $x$-$y$ and $x$-$z$ planes (b) DFT-PBEsol electronic band structure (solid black line) and its Wannier interpolation (dashed red line). The Brillouin zone is also shown (c) Detail of the electronic band structure and (d) DOS in a small region near the VHS. The energy at the VHS is set to zero.}
\label{fig:1}
\end{figure}

Phosphorene is a semiconductor with a relatively large bandgap that is underestimated (0.72~eV) at the PBEsol level (a well-known deficiency of local and semilocal DFT functionals\cite{perdew1983}), and enlarged at $1.6-2.0$~eV when $GW$ corrections\cite{hybertsen1986,hedin1965} are included.\cite{rudenko2014,Ziletti2015,tran2014}
The electronic band structure of phosphorene, calculated with the PBEsol functional, is shown in Fig. \ref{fig:1}(b).
The black solid line represents a standard plane wave DFT calculation, while the red dotted line is the band structure obtained through a Wannier interpolation. Our Wannier interpolation is very accurate, over a broad range of energies. In particular, at $\Gamma $ point, the DFT band structure and the Wannier interpolation differ by less than 10$^{-5}$~eV.

In agreement with recent studies,\cite{rodin2014,PLi2014,tran2014} we find that the top of the valence band is slightly away from the $\Gamma $ point for the LDA, PBE and PBEsol functionals.
Using these three DFT functionals, we consistently find that the top of the valence band is displaced from $\Gamma $ along the $\Gamma $-X direction, which is the direction along the phosphorene zigzag ridges [see Fig. \ref{fig:1}(a)].
The detailed symmetry analysis presented in Ref. \onlinecite{PLi2014} attributes the absence of direct bandgap to the counteracting effects (in the $\mathbf{k}\cdot \mathbf{\hat{p}}$ approximation\cite{CKittel1987}) of states of different symmetries on the valence band around the zone center.

To further validate these results, we have carried out calculations using the screened hybrid HSE06 functional and the $GW$ approximation, which are known to improve the description provided by local or semilocal DFT functionals (such as LDA,  PBE and PBEsol), not only regarding bandgaps, but also concerning the band dispersion in semiconductors and insulators.\cite{janesko2009,fuchs2007,henderson2011,aryasetiawan1992,kralik1998,yanagisawa2014,yanagisawa2013}

To quantitatively characterize the valence band maximum, we define \textbf{k}$_{\rm max}$$\equiv$$(k_{\max},0)$ as the wavector at which the valence band has a maximum, and $E_{\rm max}$ as the difference in energy between the valence band maximum and the value of the valence band at the $\Gamma$ point.
The results obtained with various computational methods and different strains are reported in Table \ref{table:vb-top}. As mentioned before, in absence of strain, LDA, PBE and PBEsol gives a valence band top slightly away from the $\Gamma$ point, with $E_{\rm max}$ ranging approximately from 1 to 12~meV. With the HSE06 functional, the valence band top is slightly misplaced from $\Gamma$; however, the calculated value of $E_{\rm max}$ is so small (0.05~meV) that it can be considered zero. Also the $GW$ method - in the absence of strain - predict phosphorene to be a direct bandgap semiconductor.

We then apply strain along the $x$-direction, changing the lattice parameter $a_x$ to be $a_x (1 + x_{\rm str})$, where a positive (negative) value of $x_{\rm str}$ indicates tensile (compressive) strain. The results are shown in Table \ref{table:vb-top}. Application of compressive strain moves the valence band top away from the $\Gamma$ point in all computational methods. In particular, with the HSE06 functional and the $GW$ method, the top of the valence band is displaced from the $\Gamma$ point with a 2\% strain; larger strains monotonically increase both $k_{\rm max}$ and $E_{\rm max}$ with $E_{\rm max}$$\sim$ 54~meV for a strain of 8\% according to the $GW$ method. In contrast, tensile strain moves the top of the valence band towards the $\Gamma$ point, and eventually removes the VHS singularity.

\begin{table*}\centering
\ra{1.3}
\begin{tabular}{@{}r c rcr c rcr c rcr c rcr c rcr c rcr@{}}\hline \hline
& \phantom{abc}& \multicolumn{3}{c}{LDA} & \phantom{abc}& \multicolumn{3}{c}{PBE} & \phantom{abc}& \multicolumn{3}{c}{PBEsol} & \phantom{abc}&\multicolumn{3}{c}{HSE06} & \phantom{abc}& \multicolumn{3}{c}{$GW$}\\
\cline{3-5} \cline{7-9} \cline{11-13} \cline{15-17} \cline{19-21} 
&&  \multicolumn{1}{c}{$k_{\rm max}$} & \phantom{i} & \multicolumn{1}{c}{E$_{\rm max}$}  &&  \multicolumn{1}{c}{$k_{\rm max}$} & \phantom{i} & \multicolumn{1}{c}{E$_{\rm max}$}  &&  \multicolumn{1}{c}{$k_{\rm max}$} & \phantom{i} & \multicolumn{1}{c}{E$_{\rm max}$}  &&  \multicolumn{1}{c}{$k_{\rm max}$} & \phantom{i} & \multicolumn{1}{c}{E$_{\rm max}$}  &&  \multicolumn{1}{c}{$k_{\rm max}$} & \phantom{i} & \multicolumn{1}{c}{E$_{\rm max}$} \\
&&  (A$^{-1}$) && (meV)  &&  (A$^{-1}$) && (meV)   &&  (A$^{-1}$) && (meV)   &&  (A$^{-1}$) && (meV)  &&  (A$^{-1}$) && (meV)  \\\hline
$x_{\rm str}=-8$\%			&& 0.256 && 79.4 	&& 0.275 && 62.3 	&& 0.260 && 73.3 && 0.271 &&	54.4	&&0.269 &&53.8\\
$x_{\rm str}=-6$\%			&& 0.212 && 51.9 	&& 0.213 && 32.9 	&& 0.211 && 44.9 && 0.168 &&	28.9	&&0.206 &&25.5\\
$x_{\rm str}=-4$\%			&& 0.175 && 33.1 	&& 0.161 && 15.1 	&& 0.175 && 27.8 && 0.129 &&	13.0	&&0.147 &&8.9\\
$x_{\rm str}=-2$\%			&& 0.145 && 20.2 	&& 0.113 &&  5.1 	&& 0.136 && 14.0 && 0.092 &&	3.6	&&0.083 &&1.1\\
$x_{\rm str}=\phantom{+}0$\%  	&& 0.119 && 11.6 	&& 0.063 && 0.7 	&& 0.105 && 6.6   && 0.029 &&	0.05	&&0.000 &&0.0\\
$x_{\rm str}=+4$\%  			&& 0.067 &&  1.9 	&& 0.000 && 0.0 	&& 0.033 && 0.1   && 0.000 &&	0.0	&&0.000 &&0.0\\
\hline \hline
\end{tabular}
\caption{Position of the valence band maximum, $k_{\rm max}$, and its energy, $E_{\rm max}$, relative to the $\Gamma$ point [see Fig.\ref{fig:1}(c)] calculated with different computational methods and various strains, $x_{\rm str}$, along the $x$-direction of the phosphorene lattice. When $k_{\rm max}$=0 (and thus $E_{\rm max}$=0) the top of the valence band coincides with the $\Gamma$ point, and therefore the VHS is not present.}
\label{table:vb-top}
\end{table*}

Having established that a VHS near the phosphorene Fermi energy is either present or can be strain-induced using a wide range of electronic structure descriptions, hereafter we consider as an explanatory example the case of the PBEsol functional. Other functionals and the $GW$ method are expected to yield similar general results.   

A magnification of the valence band maximum is shown in Fig. \ref{fig:1}(c). 
From Fig. \ref{fig:1}(c), we notice that the valence band has a saddle point at $\Gamma $. In the reciprocal space neighbourhood of this point, the principal curvature is electron-like along the $\Gamma $-X path [from the left in Fig. \ref{fig:1}(c)], while it is hole-like in the $\Gamma $-Y path [from the right in Fig. \ref{fig:1}(c)]. Thus, at the $\Gamma $ point there is a crossover from electron-like to hole-like conduction that originates at the VHS. The DOS, calculated on an ultrafine grid of $\sim$1.3 billion $k$-points, is shown in Fig. \ref{fig:1}(d). 
It exhibits a divergent behaviour at the energy position of the VHS, as expected for a 2D lattice. In contrast to the saddle point behaviour at $\Gamma$, the valence band has a maximum at ${\bf\rm k}_{\max}$ and therefore the DOS shows a step-like drop to zero at this point.\cite{grosso2000}  

\begin{figure}[htb]
\centering
    \includegraphics*[trim=12pt 140pt 16pt 140pt, width=8.6cm]{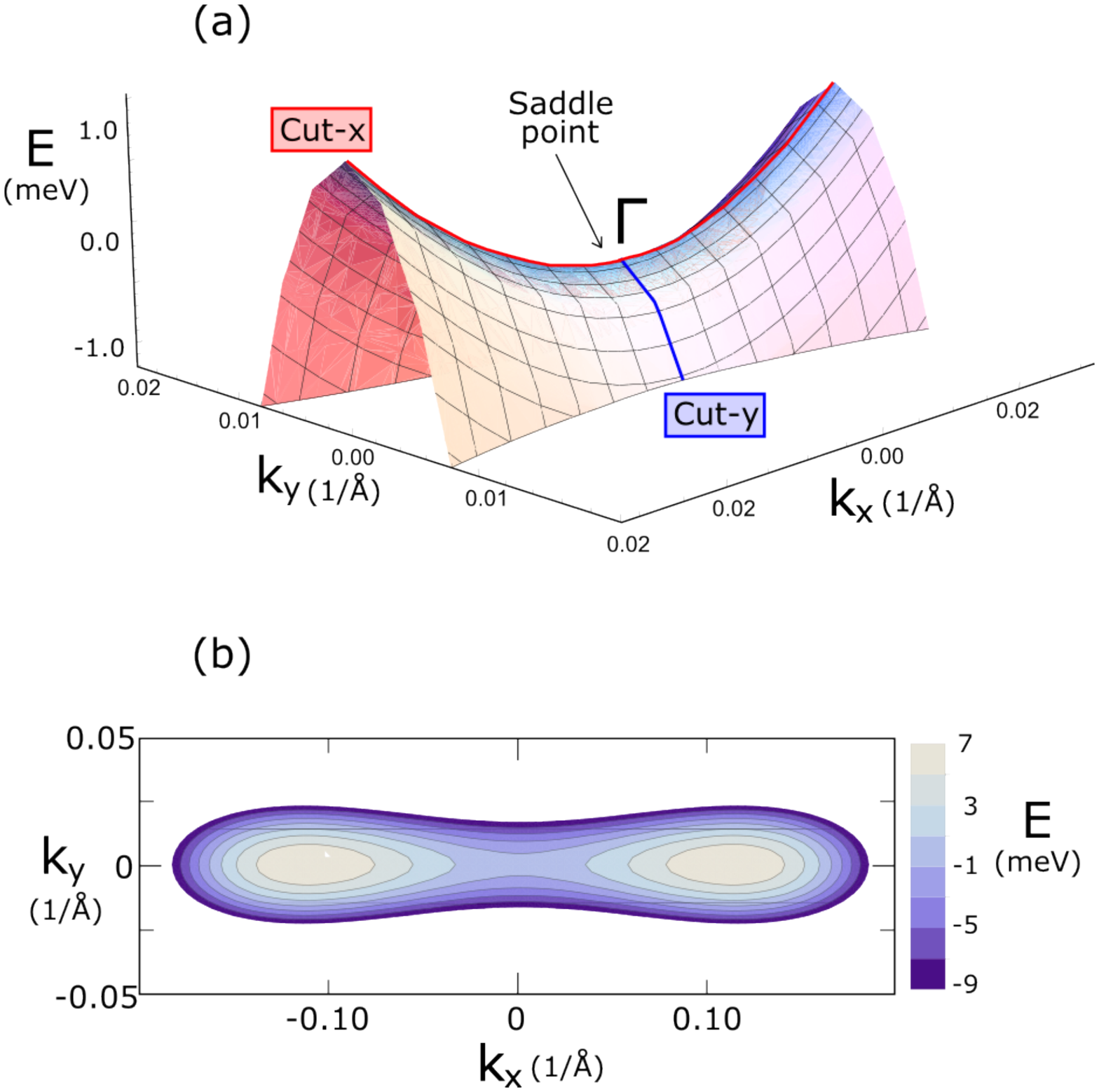}
 \caption{\small (Color online) (a) 3D plot of the phosphorene valence band around the VHS ($\Gamma $ point) (PBEsol functional). Cut-x(y) indicates the $\Gamma $-X(Y) path used in the band structure calculation. (b) 2D contour plot of the valence band in a larger region around $\Gamma $. The contour lines are drawn at 2~meV intervals. The energy at the VHS is set to zero in both plots.}
\label{fig:2}
\end{figure}

Three dimensional (3D) and 2D plots of the phosphorene valence band in the neighbourhood of the VHS ($\Gamma $ point) are depicted in Fig. \ref{fig:2}(a) and  \ref{fig:2}(b), respectively.
The electron-like dispersion along the $\Gamma $-X path [Cut-x in Fig. \ref{fig:2}(a)] and the hole-like dispersion in the $\Gamma $-Y path [Cut-y in Fig. \ref{fig:2}(a)] are evident. 
The VHS has indeed the topology of a 3D-saddle point.
Moreover, the valence band is anisotropic at the $\Gamma $ point, with strong dispersion along $\Gamma $-Y [armchair direction, see Fig. \ref{fig:1}(a)], while it is nearly flat on $\Gamma $-X [zigzag direction, see Fig. \ref{fig:1}(a)].
The large difference in magnitude of the effective masses along the two directions gives to the VHS an extended structure, as shown in Fig. \ref{fig:2}(b).
From a fitting of the local curvature of the valence band around the $\Gamma $ point, we obtain $m_x/m_y$$\sim$27, where $m_{x}$ and $m_{y}$ are the effective masses on the $\Gamma $-X and $\Gamma $-Y path, respectively.

In the limit of infinite mass in one direction (i.e. flat band in one direction, vanishing curvature) the saddle point becomes extended, giving rise to a so-called extended VHS (EVHS).
EVHSs have been experimentally observed in doped graphene\cite{mcchesney2010} and in some layered cuprate superconductors.\cite{AAbrikosov1993,KGofron1994,APiriou2011}
In 2D materials, the DOS is known to diverge logarithmically at the VHS, while in an EVHS the energy dispersion is quasi-one dimensional, and the DOS has a much stronger square-root divergence.\cite{grosso2000}
Therefore, due to the anisotropy of the phosphorene band structure, the VHS has an extended character that might amplify its effects on the material properties.

\subsection{Ferromagnetic instability}
\label{subsec:magnetism}
As mentioned in the introduction, the presence of a VHS at the Fermi energy can create ferromagnetic, antiferromagnetic or superconducting instabilities. 
In contrast to cuprates where the VHS points are at $(\pi ,0)$ and $(0,\pi )$, in phosphorene the VHS point is at $\Gamma $ and therefore we can exclude antiferromagnetism since no inter-VHS scattering can induce this instability.
Furthermore, for highly anisotropic masses (see Sec. \ref{subsec:van-hove}), $m_{x}/m_{y} \gg 1$, similar to the $t-t^{\prime}$ Hubbard model with large $t^{\prime }/t$ ($>0.276$),
the ferromagnetic instability will win over other instabilities. \cite%
{HQLin1987,Alvarez1998,Honerkamp2001} As a result, we can omit also superconductivity and consider only
ferromagnetism. 

The extremely fine structure of the VHS in phosphorene requires a very high resolution calculation of the band structure.
To make the calculation accessible, instead of using the band structure from the Wannier
interpolation, we approximate it here by an analytic single-band model.
Consistent with Fig. \ref{fig:1}(c) and Fig. \ref{fig:2}, the low-energy physics in the neighbourhood of the VHS can be described by
\begin{equation}
E\left(k_x, k_y \right)=\frac{1}{2}\alpha  k_{x}^{2}-\frac{1}{4}\beta  k_{x}^{4}-\frac{1}{2}\alpha^{\prime}  k_{y}^{2},
\label{HVHS}
\end{equation}%
which characterizes the saddle point at $\Gamma  $ (opposite signed band masses
along $k_{x}$ and $k_{y}$) and band inflection along $k_{x}$.
As in the previous section, the VHS energy at $\Gamma $, $E_{\mathrm{VHS}}$, is set to zero
while the band maximum at $k_{\max }$ is $E_{\max }$. 
To fit the DFT-PBEsol calculations, the band parameters follow the relations: $\alpha^{\prime} /\alpha $=$m_{x}/m_{y}$=27.02%
, $\sqrt{\alpha /\beta }$=$\left\vert {k}_{\max }\right\vert$=0.104~\AA$^{-1}$ and %
$\alpha ^{2}/4\beta $=$E_{\max }$=6.6~meV. 

This simple model captures the energy dispersion behaviour near the VHS and, 
since the parameters of the model are determined directly from the DFT calculation, 
it allows us to investigate the magnetic instability quantitatively. 

Obviously, a ferromagnetic instability can take place only in metallic or semimetallic systems, and therefore some amount of doping is required for phosphorene to exhibit metallic behaviour. 
To study the effects arising from the presence of the VHS, hereafter we thus assume the Fermi energy to be exactly at the VHS, $E_{\mathrm{VHS}}=0$, unless otherwise stated. In the case of the PBEsol functional, the amount of (hole) doping necessary to reach the VHS (from the top of the valence band $E_{\rm max}$) is found to be approximately 4.2$\times$10$^{-3}$ electrons per unit cell, corresponding to a doping concentration of 1.4$\times$10$^{12}$~cm$^{-2}$ for each spin.

Given the energy dispersion in Eq. (\ref{HVHS}), it is possible to derive an exact analytical expression for the DOS, $N(E)$ (see Appendix  \ref{sec:app-dos} for a complete derivation):
\begin{align}
\label{eq:dos}
N(E) = 
\begin{cases} 
-\sqrt{\frac{2}{\beta  \alpha^{\prime} }}\frac{a_x a_y}{\pi ^{2}k_{+}}K\left(\sqrt{1-p^{-2}}\right) &\hspace{-0.5em} \text{if }E \geq 0 \\
-\sqrt{\frac{2}{\beta  \alpha^{\prime} }}\frac{a_x a_y}{\pi ^{2}\left\vert k_{-}\right\vert }\frac{1%
}{\sqrt{\left\vert p\right\vert ^{2}+1}}K\left(\sqrt{\frac{1}{1+\left\vert
p\right\vert ^{-2}}}\right) &\hspace{-0.5em} \text{if }E < 0 
\end{cases}
\end{align}
where we have defined
\begin{align}
k_{\pm }=k_{\max}\sqrt{1\pm \sqrt{1-E/E_{\mathrm{\max }}}} && 
p^{2} = \frac{k_{+}^{2}}{k_{-}^{2}}
\end{align}
and $K$($k$) is the complete elliptic integral of the first kind. The quantities $a_x$ and $a_y$ are the phosphorene lattice parameters, as defined in Sec. \ref{sec:comp-details}.

Since we are mainly interested in the behaviour of the DOS at the VHS, we take the limit $E$$\rightarrow$0 in Eq. (\ref{eq:dos}) to obtain (see Appendix  \ref{sec:app-dos})
\begin{align}
\label{eq:dos-vhs}
N(E\rightarrow0_{\pm}) =\frac{a_{x}a_{y}}{2\pi ^{2}\sqrt{\alpha \alpha^{\prime} }}\left[ \ln \left( E_{\mathrm{\max }}%
/E \right)+\mathcal{O}(1)\right] 
\end{align}

The model of Eq. (\ref{HVHS}) therefore exhibits a logarithmically divergent DOS
and it is proportional to the geometric mean mass, $\sqrt{m_x m_y} \propto 1/\sqrt{\alpha \alpha^{\prime}}$, so in the limit of small energies we can approximate the DOS as
\begin{align}
\label{eq:NE0}
N(E) \approx N_{0}\ln \left(\Lambda/E\right)
\end{align}%
where $N_{0}$=$\frac{a _{x}a_{y}}{2\pi ^{2}\sqrt{\alpha  \alpha^{\prime} }}$=0.0588~eV$^{-1}$, and 
$\Lambda $ is an energy cutoff of the order of $E_{\max }$.
In this approximation, the DOS only includes contributions from states around $\Gamma  $. 
This is fully justified since the behaviour of the DOS near the VHS is obviously governed by its divergence at $E$=0 ($\Gamma $ point), and therefore finite (not diverging) contributions from other regions in the Brillouin zone can be neglected. 

\begin{figure}[tbp]
\centering
\includegraphics*[trim=30pt 70pt 30pt 60pt, width=8.6cm]{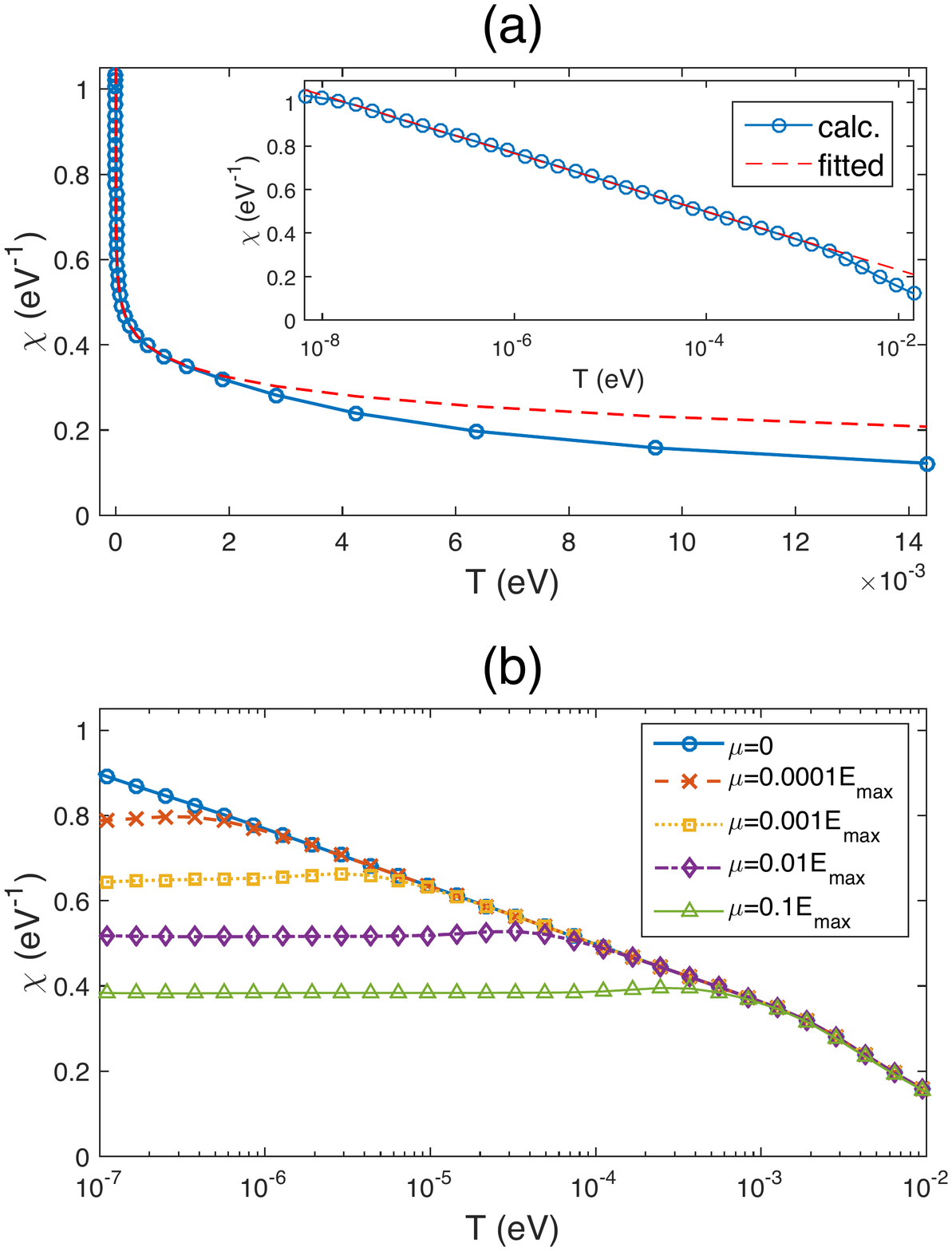}\newline
\caption{ {\small (Color online) (a) Temperature dependence of the bare spin susceptibility $\protect%
\chi $ calculated directly from Eq. (\ref{X0A}) (blue dots) or approximated with the logarithmic divergence in Eq. (\ref{X0}) (dashed red line).  
The low-temperature behaviour for $T$$<$$T^{\ast }$$\sim$1.5~
meV is seen in the inset to follow the logarithmic law. The dashed red line
is fitted to Eq. (\ref{X0}) with $\omega_D=$0.5004~eV. The Fermi level is set to $E_{\mathrm{VHS}}$=0.
(b) Spin susceptibilities $\chi$ at different Fermi energies $\mu   $. Away from the
VHS, the low-temperature logarithmic behaviour stops at $T\approx \mu  
$ and turns into Pauli susceptibility.}}
\label{fig:3}
\end{figure}

With the logarithmic DOS one can derive (see Appendix  \ref{sec:app-sus})
an expression for the bare spin susceptibility that shows a dependence on the logarithm of the inverse temperature,
\begin{align}
\protect \chi (T) =& - \int_{\rm BZ} \frac{d^2 k}{\left( 2 \pi \right)^2}\frac{\partial n_{\rm F} \left(E_k \right)}{\partial E_k} \label{X0A}\\
\approx& N_{0}\ln (\omega _{D}/T)  \label{X0}
\end{align}%
where $n_{\rm F}$ is the Fermi distribution, $E_k$ are the energy (Kohn-Sham) eigenvalues, and $\omega _{D}$ is a fitting constant of the order of $E_{\rm max}$. We also set the Boltzmann constant k$_{\rm B}$ to unity.
The logarithmic divergence at low temperatures [Eq. (\ref{X0})] is confirmed by explicit calculation of the spin susceptibility using Eq. (\ref{X0A}), as shown in Fig. \ref{fig:3}(a). 

Notably, the logarithmic behavior is present only when the temperature is lower than $T^{\ast }$, which is defined as the temperature above which the spin susceptibility starts deviating from the logarithmic behaviour. Thus, above $T^{\ast }\sim$17~K (corresponding to an energy scale of 1.5~meV), we observe deviation of the susceptibility from the logarithmic law as seen in the insert of Fig. \ref{fig:3}(a). This temperature $T^{\ast }$ is related to the energy scale $E_{\max }$.
We have checked this apparent relationship between $T^{\ast }$ and $E_{\max }$ by comparing susceptibilities for different band parameters
$\beta $ (and thus $E_{\max }$) in Eq. (\ref{HVHS}), and we indeed see proportionality
between $T^{\ast }$ and $E_{\max }$ (not shown). 
It is also found that the susceptibility increases with $E_{\rm max}$ as expected from the energy cutoff and fitting constant dependence in Eqs.(\ref{eq:NE0}) and (\ref{X0}) respectively.

Next, we examine the effect of doping on the ferromagnetic instability, considering various chemical potential shifts $\mu  $.
The susceptibility was calculated numerically and the results are presented in Fig. \ref{fig:3}(b), in which the Fermi energy is shifted to $\mu   $ above $%
E_{\mathrm{VHS}}$ (the energy of the VHS, or $\Gamma$). 
Fig. \ref{fig:3}(b) shows that even away from the VHS
point, the logarithmic-$T$ behaviour of the susceptibility is still preserved for $T$$<$$T^{\ast }$.
However, for each value of $\mu $, we see that the logarithmic increase of $\chi $ with decreasing $T$ stops at $\mu   $, below which the susceptibility become constant suggesting a transition to Pauli paramagnetism at low temperatures. 
This different behaviour for large and small $T$ (with respect to $\mu  $) can thus be understood directly from the expression of the bare spin susceptibility as outlined in Appendix B.

It is in fact possible to obtain analytical estimates for $\chi$ in both regimes (please refer to Appendix  \ref{sec:app-sus} for a complete derivation).
For $T$$\gg$$\mu  $, the susceptibility has the form
\begin{align}
\chi \left( T \gg \mu   \right) &\approx N_{0}\ln \left(\omega _{D}/T\right)\cosh ^{-2}\left(\mu   /2T\right).
\label{eq:chi_mu-bigT}
\end{align}
We observe the logarithmic-$T$ behaviour, typical of Liftshitz phase transitions. Moreover, for $\mu  /T$$\rightarrow$0, $\cosh ^{-2}(\mu  /2T)$$\rightarrow$1 and therefore, in this limit, $\chi$ for the doped system has precisely the same behaviour as in the undoped case, as confirmed by the numerical results presented in Fig. \ref{fig:3}(b).
In contrast, for $T$$\ll$$\mu  $ the susceptibility is found to be independent of $T$:
\begin{align}
\chi \left( T \ll \mu   \right) &\approx N_{0}\ln (\bar{\Lambda}/ \mu  )
\label{eq:chi_mu-smallT} 
\end{align}
where $\bar{\Lambda}$ is an energy cutoff $\bar{\Lambda}$$<$$E_{\max}$. 
This saturation of $\chi$ agrees well with the numerical results presented in Fig. \ref{fig:3}(b), and it originates from the infrared cutoff of the excitations due to the shifted thermal distribution (see Appendix  \ref{sec:app-sus}).

Now we estimate the ferromagnetic transition temperature. Let us assume a
Hubbard interaction of strength $U$ between intra-orbital spins. According
to the Stoner criterion,\cite{ECStoner1939} the magnetic transition occurs when $U_{\nu 
}\chi (T)$=1. Here $U_{\nu  }$, which is defined by $U$ times the average
weight $W_{\nu}$ at the Fermi energy for a particular orbital $\nu $, is regarded
as the effective interaction of orbital $\nu  $. 
From the Stoner criterion, the critical temperature follows the BCS form,\cite{JBardeen1957}
\begin{equation}
\label{eq:bcs}
T_{c}=\omega _{D}\exp(-1/N_{0}U_{\nu  })
\end{equation}%
where the geometric mean mass, appearing in $N_0$ as outlined above, determines the DOS at the Fermi energy. 
With this BCS formula, one can obtain the magnetization
directly. 


The effective interaction $V_{\text{eff}} $ can be evaluated using the Kohn-Sham orbitals from the DFT calculation.
Let us define orbital operators $\psi_m$ and band operators $\phi_\nu$. 
The relation between them is a unitary transformation
$\psi _{m}(\mathbf{k})$=$\sum_\nu A_{m, \nu}(\mathbf{k}) \phi_\nu (\mathbf{k})$
where $A(\mathbf{k})$ is the unitary matrix that diagonalizes the Bloch Hamiltonian.
The Hubbard onsite (intra-orbital) interaction is
\begin{align}
H_U = U\sum_{\mathbf{R},m}\psi _{m\uparrow }^{\dag }(\mathbf{R})\psi_{m\uparrow }(\mathbf{R})\psi _{m\downarrow }^{\dag }(\mathbf{R})\psi_{m\downarrow }(\mathbf{R}),
\end{align}
where $\mathbf{R}$ is the real space lattice vector and $\psi_{m \uparrow,\downarrow}$ are the Kohn-Sham spin-orbitals.

Since only the valence band (VB) is included in our low-energy model [Eq. (\ref{HVHS})], we include only the intra-band scattering terms from the Hubbard model. Moreover, at $T$=0, only states from the Fermi surface contribute to the susceptibility. After these considerations, the effective interaction $V_{\text{eff}}$ for $\nu $=VB is
\begin{align}
V_{\text{eff}}\approx U_{\nu }\sum_{\mathbf{k}_1,\mathbf{k}_2,\mathbf{k}_3}\phi _{\nu \uparrow }^{\dag }(\mathbf{k}_1)\phi _{\nu \uparrow }(\mathbf{k}_2)\phi _{\nu \downarrow }^{\dag }(\mathbf{k}_3)\phi _{\nu \downarrow }(\mathbf{k}_1+\mathbf{k}_3-\mathbf{k}_2),
\end{align}
where the momenta \textbf{k}$_1$, \textbf{k}$_2$ and  \textbf{k}$_3$ are in the neighbourhood of the Fermi surface and the interaction strength
\begin{equation}
U_{\nu }=UW_{\nu }=U\left\langle \sum_{m}\left\vert A_{m,\nu }(\mathbf{k})\right\vert ^{4}\right\rangle _{\text{FS} }
\end{equation}%
is averaged on the Fermi surface in the sense that the momentum
dependence can be neglected since the Fermi surface around the VHS is small.

From the DFT results, we obtain an average weight, $W_\nu $, for the contributing orbitals of about 0.2. This orbital weight significantly reduces the critical temperature. For
example, using the criterion $U_{\nu }\chi (T_c)$=0.8 and at $U$=4~eV, the critical temperature $T_{c}$ for ferromagnetism is
only about 4~$\mu   $K. 

Doping can destroy ferromagnetism even at zero
temperature when $N_{0}U_{\nu  }\ln (\omega _{D}/\mu   )$$<$1. Although
inter-orbital interactions might slightly enhance $T_{c}$, the Stoner criterion 
applied to the bare susceptibility typically overestimates the critical temperature
since particle-particle correlations would give large corrections to the
self-energy.\cite{Katanin2003,Katanin2011,Irkhin2001} As a result, this
ferromagnetic state would be difficult to reach. 

\subsection{Effect of strain on the van Hove singularity and on the critical temperature}
\label{subsec:strain}
Strain can have a large effect on phosphorene's pliable waved structure, and therefore it represents a natural way to tune the band parameters of the VHS, in order to increase $T_c$.
In particular, from Eq. (\ref{eq:bcs}) we notice that, for a fixed effective interaction, the critical temperature can be varied in two ways.
One way is to increase $\omega _{D}$ (or equivalently $E_{\rm max}$, see Sec. \ref{subsec:magnetism}), which will cause a linear increase in $T_c$.
The second and more prominent way is to increase $N_{0}$, which will result in an exponential increase in $T_{c}$. This can be accomplished, for instance, by reducing the dispersion in both $x$ and $y$ directions near the $\Gamma $ point [see Eqs. (\ref{HVHS}) and  (\ref{eq:NE0})].

We find that strain along the armchair direction [$y$-axis in Fig. \ref{fig:1}(a)] does not alter significantly the critical temperature, with a modest ninefold increase in $T_c$ ($\sim$36 $\mu  $K) for a tensile strain of 3\%. 

The situation, however, is significantly different for strain along the zigzag ridge direction [$x$-axis in Fig. \ref{fig:1}(a)]. The most relevant quantities for representative $x$-strain values are listed on Table  \ref{table:param-vhs}.

\begin{table}[htb]
\caption{Parameters related to the VHS for different strains on the zigzag direction [$x$-axis in Fig.\ref{fig:1}(a)]. For $T$$<$$T^{\ast}$, the susceptibility follows the logarithmic temperature dependence of Eq. (\ref{X0}). The PBEsol functional is used.}
\centering 
\begin{tabular}{l @{\hspace{1em}} | c@{\hspace{1em}} | c@{\hspace{0em}} | c@{\hspace{1em}} }
\hline \hline
& \multicolumn{1}{c|}{$x_{\rm str}=-4\%$} & \multicolumn{1}{c|}{$x_{\rm str}=0\%$} & \multicolumn{1}{c}{$x_{\rm str}=+4\%$} \\
\hline
\multicolumn{1}{l|}{\textbf{k}$_{\rm max}$ (1/\AA)}& \multicolumn{1}{c|}{(0.175,0)} & \multicolumn{1}{c|}{(0.105,0)} & \multicolumn{1}{c}{(0.033,0)}\\
\multicolumn{1}{l|}{$E_{\rm max}$ (meV)} & \multicolumn{1}{c|}{27.8} & \multicolumn{1}{c|}{6.6} & \multicolumn{1}{c}{0.1}\\
\multicolumn{1}{l|}{$N_0$ (eV$^{-1}$)} & \multicolumn{1}{c|}{0.0374} & \multicolumn{1}{c|}{0.0588} & \multicolumn{1}{c}{0.1909}\\
\multicolumn{1}{l|}{$T^{\ast}$ (K)} & \multicolumn{1}{c|}{97} & \multicolumn{1}{c|}{17} & \multicolumn{1}{c}{0.23}\\
\multicolumn{1}{l|}{$T_{c}$ (K)} & \multicolumn{1}{c|}{5$\times$10$^{-7}$} & \multicolumn{1}{c|}{4$\times$10$^{-6}$} & \multicolumn{1}{c}{5$\times$10$^{-2}$}\\
\hline\hline 
\end{tabular}
\label{table:param-vhs}
\end{table}

\begin{figure}[htb]
\centering
    \includegraphics*[trim=0pt 0pt 0pt 0pt, width=8.4cm]{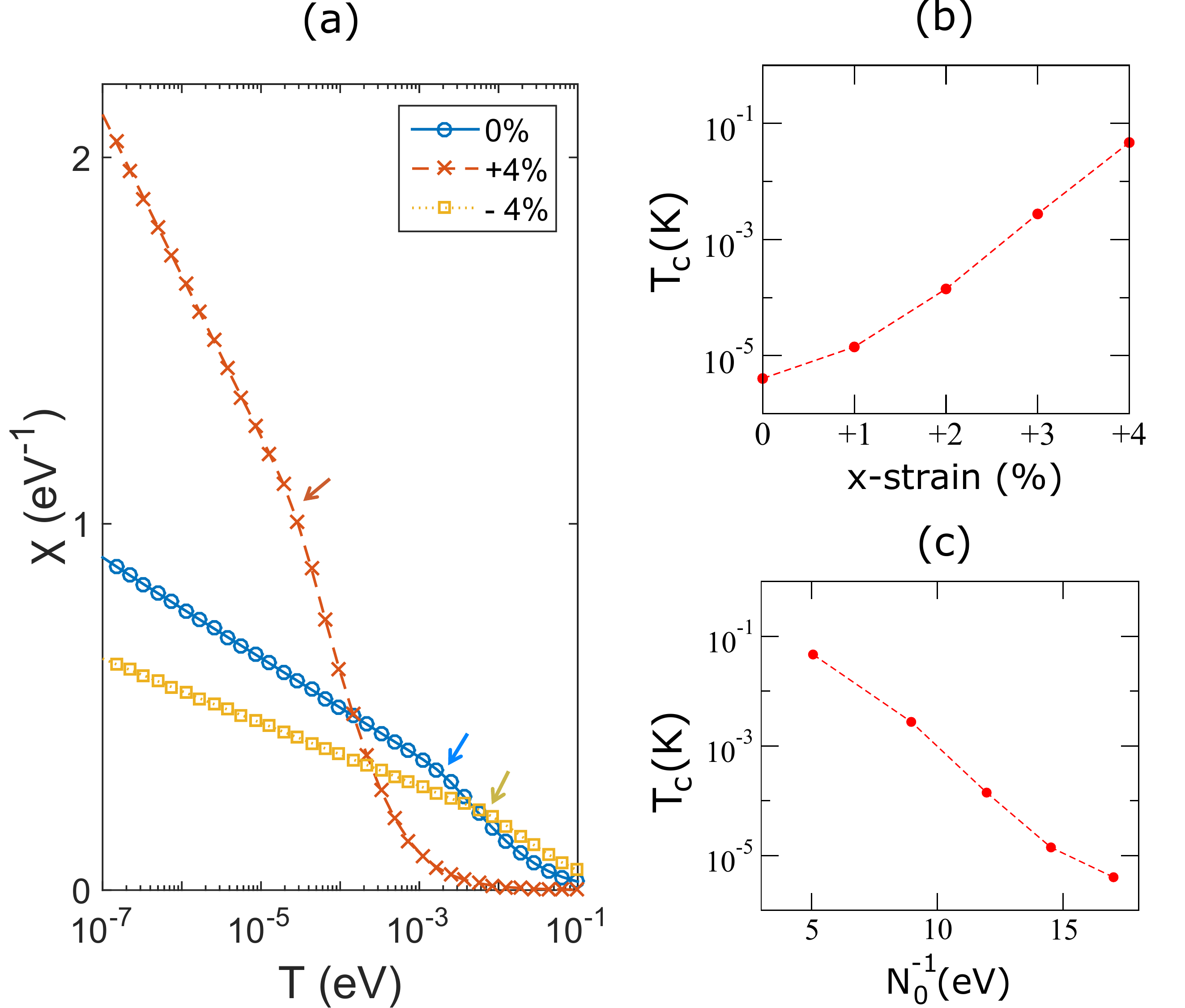}
 \caption{\small (Color online) Effect of zigzag ridge [$x$-axis in Fig.\ref{fig:1}(a)] strain on the VHS (a) Temperature dependence of the bare spin susceptibility $\protect \chi$ for different strains. The arrows indicate the temperature $T^{\ast}$ at which the susceptibility starts to deviate from the logarithmic behaviour. (b) Critical temperature $T_c$ as a function of strain. (c) $T_c$ versus $N_0^{-1}$. The behaviour of $T_c$ still follows the exponential law of Eq. (\ref{eq:bcs}) even if strain is applied. }
\label{fig:4}
\end{figure}

Compressive $x$-strain of 4\% slightly reduces $N_0$ (see Table \ref{table:param-vhs}), and leads to a decrease in $T_c$ to only 0.5~$\mu  $K. In contrast to this $N_0$, we see from Table \ref{table:param-vhs} that $E_{\max}$ increases with this compressive $x$-strain. Due to the proportionality between $E_{\max}$ and $T^{\ast}$, the spin susceptibility starts to follow the logarithmic-$T$ behaviour - the signature of the VHS - at higher temperatures than the unstrained case. For example, we see that a compressive $x$-strain of 4\% leads to $T^{\ast}$ of about 97~K (8.4~meV) [see Fig. \ref{fig:4}(a), yellow squares].
Due to the relatively high temperatures involved, the logarithmic-$T$ behaviour in the spin susceptibility could in principle be observable experimentally, thus providing compelling evidence for the presence of the VHS.

In contrast, tensile $x$-strain has the opposite effect on the band parameters: while $E_{\max}$ diminishes, $N_0$ is greatly enhanced. Notably, the critical temperature exhibits an exponential dependence on tensile strain, as depicted in Fig. \ref{fig:4}(b). For a 4\% strain, the critical temperature is about 0.05~K. Even though this corresponds to a 10$^{4}$-fold increase in $T_c$ with respect to the unstrained case, this magnetic state will hardly be seen experimentally due to the very low temperatures required. 
We also observe that the logarithmic divergence becomes the dominant contribution at around $T^{\ast }$$\sim$0.23~K (0.002~meV) [Fig. \ref{fig:4}(a), orange crosses], a much lower value compared to the unstained case resulting from the flattening of the valence band (and consequently diminished $E_{\max}$), caused by the applied stress.

The physics behind the strain-dependence of the VHS is simple. The act of stretching will decrease the hoppings between phosphorus atomic orbitals, thus reducing the bandwidth. As the dispersion decreases, we expect that $\alpha$ and $\alpha^{\prime}$ in Eq. (\ref{HVHS}) become smaller and hence the density of states $N_0$ increases. The act of compressing will show the opposite trend. Because the band inflection is along $k_x$, $x$-strain have a larger effect on $\alpha$ compared to $y$-strain, explaining our findings.

Finally, we observe that the critical temperature still follows the exponential law of Eq. (\ref{eq:bcs}), even when strain is applied, as shown in Fig. \ref{fig:4}(c). For higher stress, $T_c$ could deviate from Eq. (\ref{eq:bcs}), since $E_{\max}$ diminishes ($\sim$10$^{-4}$~eV for a 4 \% tensile strain) and therefore narrows the VHS divergence, limiting the increase of the critical temperature.

\section{Conclusions}

We have used Wannier function-based interpolation techniques to investigate the VHS at the $\Gamma $ point near the phosphorene Fermi energy with more than a billion $k$-points. Thanks to this extreme resolution, we are able to present compelling numerical evidence for the presence of a VHS near the phosphorene Fermi energy. 
As a result of its close proximity to the valence band maximum, the VHS can be reached with a hole doping concentration on the order of 10$^{12}$ cm$^{-2}$, easily achievable by chemical doping or ionic-liquid gating.\cite{saito2015}

Furthermore, we have calculated an exact expression for the DOS near the VHS, and we have demonstrated that the spin susceptibility presents a logarithmic-$T$ behaviour, signature of the VHS, and consequent Liftshitz phase transition.

We have also shown that the critical temperature can be increased up to 0.05~K by applying a modest strain to the phosphorene pliable waved structure. 
Although this ferromagnetic state would be very difficult to reach experimentally, the logarithmic temperature behaviour of the spin susceptibility due to the presence of the VHS could be observed because it persists at higher temperatures ($T^{\ast}$$\sim$17~K for the unstrained case, and $T^{\ast}$$\sim$97~K for a 4\% tensile strain along the zigzag ridges).  

There are numerous experimental techniques able to detect the presence of VHSs.
For example, the scanning tunnelling microscope (STM) measures the tunneling differential conductance, which is proportional to the local DOS,\cite{JTersoff1985} and therefore represents an ideal tool to detect VHSs. 
This technique has been used to observe VHSs in other 2D materials like twisted multilayer graphene,\cite{GLi2010} or the cuprate superconductor Bi-2201.\cite{APiriou2011}
Furthermore, angle-resolved photoemission spectroscopy (ARPES) can detect saddle points in the single-particle energy dispersion, as employed for numerous cuprate compounds\cite{AAbrikosov1993,DSDessau1993,KGofron1994,DMKing1994} and doped graphene.\cite{mcchesney2010} 
Finally, the Knight shift\cite{Knight2007} in nuclear magnetic resonance experiments could provide evidence for the change in spin susceptibility in proximity of the VHS.

\section*{Acknowledgements}
We thank Chuang-Han Hsu for technical assistance.
A.Z. and D.F.C. acknowledge NSF grant CHE-1301157 and also an allocation of computational resources from Boston University's Office of Information Technology and Scientific Computing and Visualization.
H.L. acknowledges the Singapore National Research Foundation for the
support under NRF Award No. NRF-NRFF2013-03.


\appendix
\section{Exact derivation of the density of states}
\label{sec:app-dos}

In this section, we present an analytical derivation of the DOS for phosphorene around the VHS.
According to Eq. (\ref{HVHS}), the dispersion relation of the valence band around the $\Gamma $ point has the form 
\begin{align}
E_k=\frac{1}{2}\alpha k_{x}^{2}-\frac{1}{4}\beta k_{x}^{4}-\frac{1}{2}\alpha^{\prime} k_{y}^{2} 
\label{appendix-HVHS}
\end{align}
with $\alpha ,\beta ,\alpha^{\prime} >0$.
The valence band has its energy extreme at $E_{\mathrm{\max }}=%
\alpha ^{2}/4\beta $ when ($k_{x}$,$k_{y}$)$=(k_{\max
},0)$ and $k_{\max }=\sqrt{\alpha /\beta }$. Moreover, there is a VHS at energy $E_{\mathrm{VHS}}=0$ originating from states near $k_{x}=0$.

By definition, the DOS per spin per unit area is
\begin{align}
N(E) &=\int \frac{dk_{x}dk_{y}}{(2\pi )^{2}}\delta (E-E_{k})  \label{DOS_1}\\
&=\frac{2}{(2\pi )^{2}}\int dk_{x}\int_{k_{y}\geq 0}dk_{y}\frac{1}{%
\left\vert \partial _{k_{y}}E_{k}\right\vert }\delta (k_{y}-k_{y}^{E})
\notag \\
&=\frac{1}{\sqrt{2\alpha^{\prime} }\pi ^{2}}\int_{k_{x}\geq 0}dk_{x}\frac{1}{\sqrt{\frac{\alpha %
}{2}k_{x}^{2}-\frac{\beta }{4}k_{x}^{4}-E}}.  \notag
\end{align}%
where $k_{y}^{E}$ satisfies $E=\frac{1}{2}\alpha k_{x}^{2}-\frac{1}{4}\beta k_{x}^{4}-%
\frac{1}{2}\alpha^{\prime} \left( k_{y}^{E}\right) ^{2}$.

The integration range is limited by the fact that the square root term has
to be real. After some algebra, one can show that 
the integral range is $k\in \lbrack \max (0,k_{-}),k_{+}]$
with
\begin{align}
k_{\pm }=&\sqrt{\frac{\alpha }{\beta }\pm \sqrt{\left(\frac{\alpha }{\beta }\right)^{2}-\frac{4E}{\beta }}}
=k_{\max}\sqrt{1\pm \sqrt{1-E/E_{\mathrm{\max }}}}.
\end{align}%
Thus, if $E < 0$, $%
k_{-}\ $ is not purely real, and the lower bound is zero. Therefore, we have
\begin{align}
&\int_{0}^{k_{+}}dk_{x}\frac{1}{\sqrt{\frac{\alpha }{2}k_{x}^{2}-\frac{\beta }{4}%
k_{x}^{4}-E}}  \label{E1} \\
=&\int_{0}^{k_{+}}dk_{x}\frac{1}{\sqrt{\left( -\frac{\beta }{4}\right)
(k_{x}^{2}-k_{+}^{2})(k_{x}^{2}-k_{-}^{2})}} \notag \\
=&\frac{-2i}{\sqrt{\beta }k_{-}}\int_{0}^{1}dx\frac{1}{\sqrt{%
(1-x^{2})(1-p^{2}x^{2})}}  \notag \\
=&\frac{-2i}{\sqrt{\beta }k_{-}}F\left(\frac{\pi }{2},p\right) = \frac{-2i}{\sqrt{\beta }k_{-}}K(p).  \notag
\end{align}%

where we have defined
\begin{equation}
p^{2}\equiv \frac{k_{+}^{2}}{k_{-}^{2}}=\frac{E_{\mathrm{\max }}}{E}\left( 1+%
\sqrt{1-\frac{E}{E_{\mathrm{\max }}}}\right) ^{2}
\end{equation}%
and introduced $F(\phi ,k)$ the incomplete elliptic integral of the first
kind and $K(k)$ the complete elliptic integral of the first kind. They are
related here by $F(\frac{\pi }{2},k)=K(k)$.

On the other hand, if $E>0$ ($k_{-} > 0$), the lower bound is $k_{-}$ and we
will deal with%
\begin{align}
&\int_{k_{-}}^{k_{+}}dk_{x}\frac{1}{\sqrt{\frac{\alpha }{2}k_{x}^{2}-\frac{\beta }{4}%
k_{x}^{4}-E}}  \label{E2} \\
=&\int_{k_{-}}^{k_{+}}dk_{x}\frac{1}{\sqrt{\left( -\frac{\beta }{4}\right)
(k_{x}^{2}-k_{+}^{2})(k_{x}^{2}-k_{-}^{2})}} \notag \\
=&\frac{-2i}{\sqrt{\beta }k_{-}}\int_{p^{-1}}^{1}dx\frac{1%
}{\sqrt{(1-x^{2})(1-p^{2}x^{2})}}  \notag \\
=&\frac{-2i}{\sqrt{\beta }k_{-}}\left( \frac{-i}{p}\right) F \left( \frac{\pi }{2},%
\sqrt{1-p^{-2}} \right) \notag \\
=& \frac{-2}{\sqrt{\beta }k_{+}}K(\sqrt{1-p^{-2}}). \notag 
\end{align}%
In Eqs.  (\ref{E1}) and  (\ref{E2}), we have used the integral formulae:
\begin{widetext}
\begin{align}
\int_{0}^{u}dx\frac{1}{\sqrt{(1-x^{2})(1-p^{2}x^{2})}} &=\frac{1}{2}%
\int_{0}^{u^{2}}dz\frac{1}{\sqrt{z(1-z)(1-p^{2}z)}}=F\left(\arcsin (u),p\right) \\
\int_{u}^{1}dx\frac{1}{\sqrt{(1-x^{2})(1-p^{2}x^{2})}} &=\frac{1}{2}%
\int_{u^{2}}^{1}dz\frac{1}{ip\sqrt{z(1-z)(z-p^{-2})}}=\frac{-i}{p}F\left(\arcsin \left(%
\sqrt{\frac{1-u^{2}}{1-p^{-2}}}\right),\sqrt{1-p^{-2}}\right).
\end{align}%
\end{widetext}

For negative energies, $E<0$, 
\begin{align}
&k_{-}=i\sqrt{\sqrt{1+\left\vert
E\right\vert /E_{\mathrm{\max }}}-1}=i\left\vert k_{-}\right\vert \\ 
&p=i \sqrt{\frac{E_{\mathrm{\max }}}{\left\vert E\right\vert }}\left( 1+\sqrt{1+%
\frac{\left\vert E\right\vert }{E_{\mathrm{\max }}}}\right) =i\left\vert
p\right\vert 
\end{align}%
and therefore we will use the relation
\begin{equation}
K(ik)=\frac{1}{\sqrt{k^{2}+1}}K\left( \sqrt{\frac{k^{2}}{k^{2}+1}}\right)
\end{equation}%
in Eq.  (\ref{E1}). 

In conclusion, the density of
states for $E<0$ and $E>0$ are, respectively,
\begin{align}
N(E <0)=&\frac{1}{\sqrt{2\alpha^{\prime} }\pi ^{2}}\int_{0}^{k_{+}}dk_{x}\frac{1}{\sqrt{%
\frac{\alpha }{2}k_{x}^{2}-\frac{\beta }{4}k_{x}^{4}-E}} \\
=&-i\sqrt{\frac{2}{\beta \alpha^{\prime} }}\frac{1}{\pi ^{2}i\left\vert k_{-}\right\vert }%
K\left(i\left\vert p\right\vert \right)  \notag \\
=&-\sqrt{\frac{2}{\beta \alpha^{\prime} }}\frac{1}{\pi ^{2}\left\vert k_{-}\right\vert }\frac{1%
}{\sqrt{\left\vert p\right\vert ^{2}+1}}K\left(\sqrt{\frac{1}{1+\left\vert
p\right\vert ^{-2}}}\right)  \notag
\end{align}
and
\begin{align}
N(E >0)=&\frac{1}{\sqrt{2\alpha^{\prime} }\pi ^{2}}\int_{k_{-}}^{k_{+}}dk_{x}\frac{1}{\sqrt{\frac{\alpha }{2}k_{x}^{2}-\frac{\beta }{4}k_{x}^{4}-E}} \\
=&-\sqrt{\frac{2}{\beta \alpha^{\prime} }}\frac{1}{\pi ^{2}k_{+}}K\left(\sqrt{1-p^{-2}}\right)  \notag
\end{align}

The final step is to analyze the asymptotic behaviour of the DOS.
Since $K$ will show a logarithmical divergence when
\begin{equation}
K(k=1-\eta )\overset{\eta \rightarrow 0}{\longrightarrow }\frac{1}{2}%
\ln \left\vert \eta \right\vert +O(1),
\end{equation}%
for $E\rightarrow 0$, both the quantities $\sqrt{\frac{1}{1+\left\vert p\right\vert ^{-2}}}$
and $\sqrt{1-p^{-2}}$ approach one. By using $k_{\pm }\rightarrow \sqrt{%
2}k_{\max }$, $\left\vert k_{-}\right\vert \rightarrow k_{\max }\sqrt{\frac{%
\left\vert E\right\vert }{2E_{\mathrm{\max }}}}$, $\left\vert p\right\vert
\rightarrow 2\sqrt{\frac{E_{\mathrm{\max }}}{\left\vert E\right\vert }}$,
and $\frac{1}{\sqrt{\left\vert p\right\vert ^{2}+1}}\rightarrow \frac{1}{2}%
\sqrt{\frac{\left\vert E\right\vert }{E_{\mathrm{\max }}}}$, we obtain the DOS at the VHS as:

\begin{align}
N(E \rightarrow 0_{-})
=\frac{1}{2\pi ^{2}}\sqrt{\frac{1}{\alpha \alpha^{\prime} }}\left[ \ln \left( \frac{E_{\mathrm{\max }}%
}{\left\vert E\right\vert } \right)+O(1)\right]   
\end{align}%
and similarly,%
\begin{align}
N(E \rightarrow 0_{+}) =\frac{1}{2\pi ^{2}}\sqrt{\frac{1}{\alpha \alpha^{\prime} }}\left[ \ln \left(\frac{E_{\mathrm{\max }}%
}{E} \right)+O(1)\right] 
\end{align}
After multiplication by the unit cell area $a_x$$\times$$a_y$, we obtain the result reported in Eq. (\ref{eq:dos-vhs}). 


\section{Bare spin susceptibility at the VHS and the effect of doping}
\label{sec:app-sus}


Firstly, we derive the bare susceptibility when the Fermi energy is at $E_{%
\mathrm{VHS}}$=0 ($\mu  $=0). The spin susceptibility is given by%
\begin{align}
&\chi (T)= \frac{a_{x}a_{y}}{(2\pi )^{2}}\int d^{2}k\frac{1}{4T}\cosh ^{-2}\left(\frac{%
E _{k}}{2T}\right)  \notag \\
=&\frac{1}{4T}\int_{-\infty }^{E_{\max }}dE N(E )\cosh
^{-2}\left(\frac{E }{2T}\right)  \notag \\
\approx &\frac{1}{4T}\int_{-\Lambda }^{\Lambda }dE N_{0}\ln
\left( \frac{\bar{\Lambda}}{|E |}\right) \cosh ^{-2}\left(\frac{%
E }{2T}\right)  \notag \\
=&\frac{1}{2}N_{0}\int_{-\Lambda /2T}^{\Lambda /2T}dx \ln \left(
\frac{\bar{\Lambda}/2T}{|x|}\right) \cosh ^{-2}x  \notag \\
=&-N_{0}\int_{0}^{\Lambda /2T}dx\frac{\ln x}{\cosh ^{2}x}+ \notag \\
&+N_{0}\ln \left(\frac{%
\bar{\Lambda}}{2T}\right)\int_{0}^{\Lambda /2T}dx\cosh ^{-2}x  \notag 
\end{align}%
where we have considered only the contribution from the VHS and
the logarithmic behaviour applies when $|E |<\Lambda $ and $\bar{\Lambda}$
is another energy cutoff $\bar{\Lambda}$$<$$E_{\max }$. 

If we then consider the limit $\Lambda \gg T$ and use the following formulae:
\begin{align}
\int_{0}^{\infty }dx\frac{\ln x}{\cosh ^{2}x} =\log \frac{\pi }{4}-\gamma_e  \equiv -C=-0.8188
\end{align}
($\gamma_e $ is the Euler-Mascheroni constant) \\
\begin{align}
\int_{0}^{\Lambda /2T}dx\cosh ^{-2}x &=\tanh \frac{\Lambda }{2T}\approx 1
\end{align}
we obtain the following expression for the bare susceptibility
\begin{align}
\label{eq:X0-no-dop}
\protect \chi (T) \approx &N_{0}C+N_{0}\ln (\frac{\bar{\Lambda}}{2T})\equiv N_{0}\ln (\omega
_{D}/T)
\end{align}%
which is Eq. (\ref{X0}) in the main text.


Then, we discuss the effect of doping on the susceptibility. 
When we shift the Fermi energy from zero ($%
E_{\mathrm{VHS}}$) to $\mu   $, the VHS will change to $-\mu   $, and therefore we can replace
the DOS by $N(E )\approx N_{0}\ln \frac{\Lambda }{%
|E +\mu   |}$ for $|E +\mu   |<\Lambda $. 
The susceptibility thus becomes
\begin{align}
\label{eq:app-X0}
\chi(T) &\approx \frac{1}{4T}N_{0}\int_{-\Lambda -\mu   }^{\Lambda -\mu  
}dE \ln \left( \frac{\bar{\Lambda}}{|E +\mu   |}\right)
\cosh ^{-2}\left(\frac{E }{2T}\right) \\
&=\frac{1}{4T}N_{0}\int_{-\Lambda }^{\Lambda }dE \ln \frac{\bar{%
\Lambda}}{|E |}\cosh ^{-2}\left( \frac{E -\mu   }{2T}\right) \notag.
\end{align}%

We will now consider two regimes: $T\gg \mu   $ and $T\ll \mu  $.

Let us start with the case $T$$\gg$$\mu  $. 
By using the expansion
\begin{align}
\begin{split}
\cosh \left(\frac{E -\mu   }{2T}\right)&=\cosh \left(\frac{E }{2T}\right)\cosh \left(%
\frac{\mu   }{2T}\right)+ \\
&- \sinh \left(\frac{E }{2T}\right)\sinh \left(\frac{\mu   }{2T}\right),
\end{split}
\end{align}%
and the approximation 
\begin{align}
\cosh ^{-2}\left(\frac{E -\mu   }{2T}\right)\approx
\cosh ^{-2}\left(\frac{E }{2T}\right)\cosh ^{-2}\left(\frac{\mu   }{2T}\right),
\end{align}
the susceptibility can be written as%
\begin{align}
\chi(T \gg \mu  )&\approx \frac{1}{4T}N_{0}\int_{-\Lambda }^{\Lambda
}dE \ln \frac{\bar{\Lambda}}{|E |}\cosh ^{-2}\left(\frac{%
E }{2T}\right)\\
&\hspace{4em} \times \cosh ^{-2}\left(\frac{\mu   }{2T}\right) \notag\\
&=N_{0}\ln \left(\omega _{D}/T\right)\cosh ^{-2}\left(\frac{\mu   }{2T}\right) \notag \\
&\approx N_{0}\ln \left(\omega _{D}/T\right)
\end{align}%
and therefore, in this regime, the susceptibility is the same as the undoped case [compare to Eq. (\ref{eq:X0-no-dop})].

On the other hand, when $T$$\ll$$\mu  $, the function $\cosh ^{-2}(\frac{%
E -\mu   }{2T})$ decreases proportionally to  $\exp (\frac{\mu  
-E }{T})$ for $\left\vert E -\mu   \right\vert \gg T$.
Therefore it is a good approximation to replace $\ln \frac{\bar{\Lambda%
}}{|E |}\cosh ^{-2}(\frac{E -\mu   }{2T})$ by $\ln \frac{%
\bar{\Lambda}}{|\mu   |}\cosh ^{-2}(\frac{E -\mu   }{2T})$ in the
integrand in Eq. (\ref{eq:app-X0}).  
As a result,
\begin{align}
\chi(T\ll \mu  ) &\approx \frac{1}{4T}N_{0}\int_{-\Lambda }^{\Lambda
}dE \ln \frac{\bar{\Lambda}}{|E |}\cosh ^{-2}\left(\frac{%
E -\mu   }{2T}\right) \notag \\
\begin{split}
&=\frac{1}{4T}N_{0}\int_{0}^{\Lambda }dE \ln \frac{\bar{\Lambda}}{%
|E |}\left[ \cosh ^{-2}\left(\frac{E -\mu   }{2T}\right)\right.+\\
&\left.\vphantom{\cosh ^{-2}\left(\frac{E +\mu   }{2T}\right)} + \cosh ^{-2}\left(\frac{E +\mu   }{2T}\right) \right]  \notag 
\end{split}\\
\begin{split}
&\approx \frac{1}{4T}N_{0}\int_{0}^{\Lambda }dE \ln \frac{\bar{%
\Lambda}}{\mu   }\left[ \cosh ^{-2}\left(\frac{E -\mu   }{2T}\right)\right.+ \\
&\left. \vphantom{\cosh ^{-2}\left(\frac{E +\mu   }{2T}\right)} +\cosh ^{-2}\left(\frac{E +\mu   }{2T}\right) \right] 
\end{split}
\label{eq:supp-X0small}
\end{align}
Then, using
\begin{align}
\begin{split}
\int_{0}^{\Lambda }dE \cosh ^{-2}\left(\frac{E \pm \mu   }{2T}%
\right)=2T &\left[ \tanh \left(\frac{\Lambda \mp \mu   }{2T}\right)\pm \right.\\
\pm & \left. \vphantom{\tanh \left(\frac{%
\mu   }{2T}\right)} \tanh \left(\frac{%
\mu   }{2T}\right)\right]
\end{split}
\end{align}%
we obtain
\begin{align}
\chi(T\ll \mu  )=&\frac{1}{2}N_{0}\ln \frac{\bar{\Lambda}}{\mu   }\left[ \tanh \left( \frac{\Lambda
+\mu   }{2T}\right)+\tanh \left(\frac{\Lambda -\mu   }{2T}\right) \right]  \notag \label{X0_mu}\\
=&\frac{1}{2}N_{0}\ln \frac{\bar{\Lambda}}{\mu   }\frac{\sinh \frac{\Lambda }{%
T}}{\cosh \frac{\Lambda +\mu   }{2T}\cosh \frac{\Lambda -\mu   }{2T}}  \notag \\
\approx &N_{0}\ln \frac{\bar{\Lambda}}{\mu   }\tanh \frac{\Lambda }{T}\approx
N_{0}\ln \frac{\bar{\Lambda}}{\mu   } 
\end{align}
The final result of Eq.  (\ref{X0_mu}) indicates that, in this regime, the bare spin susceptibility is independent of temperature, as seen in Fig. \ref{fig:3}(b). This behaviour originates from the infrared cutoff of the excitations due to the thermal distribution.
Let us in fact consider the integrand in Eq. (\ref{eq:supp-X0small}):
\begin{align}
\begin{split}
\ln \frac{\bar{\Lambda}}{|E |}\left[ \cosh ^{-2}\left(\frac{E -\mu   }{2T}%
\right)+\cosh ^{-2}\left(\frac{E +\mu   }{2T}\right)\right] \\
\simeq\ln \frac{\bar{%
\Lambda}}{|E |}\cosh ^{-2}\left(\frac{E -\mu   }{2T}\right)
\end{split}
\end{align}%
Due to the chemical potential shift $\mu  $, the thermal distribution function, $\cosh ^{-2}(\frac{%
E -\mu   }{2T})$$\sim$$\exp \left[ -\frac{1}{\sqrt{2}}(\frac{%
E -\mu   }{2T})^{2}\right] $, is now centered at $E$=$\mu  $ and not at $E$=0 like in the undoped case.
Therefore, the integration around $E$=0, where the logarithmic function diverges, now makes essentially no contribution to the total integral, giving raise to the flattening of the susceptibility observed for $T$$\ll$$\mu  $.

\bibliographystyle{unsrtnat}

\end{document}